\documentclass[
aps,
prl,
amsmath,
amssymb,
aps,
superscriptaddress,
twocolumn,
]{revtex4-2}
\usepackage[T1]{fontenc} 
\usepackage{graphicx}
\usepackage{dcolumn}
\usepackage{bm}
\usepackage{hyperref}
\usepackage{xcolor}
\usepackage{tikz}
\usepackage{endnotes}

\let\footnote\endnote
\usepackage[normalem]{ulem} %makes underlining possible: \uline

\newcommand{\old}[1]{\textcolor{red}{\sout{#1}}}
\newcommand{\note}[1]{\textcolor{magenta}{[#1]}}

\begin{document}

\title{
Phase-space microscopes for quantum gases:  Imaging conjugate variables and momentum-weighted densities.}

\author{N. R. Cooper}
\affiliation{TCM Group, Cavendish Laboratory, University of Cambridge, J.J. Thomson Avenue, Cambridge, CB3 0US, United Kingdom\looseness=-1}

\author{Y. Yang}
\affiliation{TCM Group, Cavendish Laboratory, University of Cambridge, J.J. Thomson Avenue, Cambridge, CB3 0US, United Kingdom\looseness=-1}

\author{C. Weitenberg}
%\email {christof.weitenberg@tu-dortmund.de}
\affiliation{Department of Physics, TU Dortmund University, 44227 Dortmund, Germany}

\begin{abstract}

{Quantum gas microscopes offer unprecedented insights into quantum many-body states of cold atomic gases. Here we introduce concrete protocols for extending quantum gas microscopes to measure in phase space, by mapping momentum onto auxiliary degrees of freedom and using positive operator-valued measures. We distinguish between two distinct operational modes. In the Husimi-Q phase space microscope, position and momentum are jointly measured; in this mode the fundamental quantum noise  is distributed between position and momentum. Conversely, the averaged-mode phase space microscope extracts the spatial dependence of averages of the momentum density (and its moments); these averages can be retrieved with arbitrary spatial resolution. We illustrate the utility of these techniques in diverse physical settings.}

\end{abstract}

\date{\today}

\maketitle

{Quantum gas microscopes (QGMs) perform the spatially resolved detection of the individual particles in a quantum many-body system of ultracold atoms and allow for microscopic insights into quantum phases~\cite{Gross2021}. They were first realised in lattice systems~\cite{Bakr2010,Sherson2010} and later in the continuum~\cite{Holten2021,Jongh2025,Xiang2025,Yao2025}. Projective measurements of atomic positions allow access not only to density-density correlators~\cite{Endres2011}, but also to various multi-point correlators including string order~\cite{Endres2011,Hilker2017,Koepsell2019}. Recent experiments went beyond detection in the position basis by measuring local currents and kinetic energy operators along certain bonds of an optical lattice~\cite{Impertro2024, Impertro2025} as well as local phases on the lattice sites~\cite{Bruggenjurgen2024}. An effective way to overcome the optical diffraction limit is the matter-wave microscope, which magnifies the density distribution prior to optical imaging using two matter-wave lenses formed by quarter-period evolutions in harmonic traps~\cite{Asteria2021,Brandstetter2025}. This method also gives access to a matter-wave Fourier space and allows manipulation to provide an analogue of phase-contrast imaging~\cite{Murthy2019,Bruggenjurgen2024} or access to various long-range off-diagonal correlators~\cite{Weitenberg2026}. 

{Here we show how to extend these techniques to phase-space microscopes, which access atomic occupations and correlations {jointly} in both position and momentum. Despite being conjugate variables -- and therefore non-commuting -- joint measurements of position and momentum are permitted, albeit at the expense of introducing quantum noise via the measurement apparatus~\cite{Arthurs1965,Peres1995}. We shall show how this can be done for cold gases by mapping momentum (or some function of it) onto  auxiliary degrees of freedom. This allows positive operator-valued measures (POVMs) that reflect the non-commuting observables~\cite{Peres1995}.  We discuss protocols for mapping momentum space to an auxiliary dimension or to an auxiliary spin by appropriate manipulation in matter-wave Fourier space.}  We further show that the approach can be adapted to allow measurements of {\it averages} of the momentum density (or its moments). Such observables can, in principle, be measured with arbitrary spatial resolution. 
We illustrate the utility of phase-space microscopes for a variety of physical settings.

We first introduce the phase-space microscope for a single particle of mass $M$ moving in a continuous system in one dimension, with system co-ordinate $x$, and using the $z$-dimension as the auxiliary space (Fig.~\ref{fig:1}). 
\begin{figure}
    \centering
    \includegraphics[width=0.95\linewidth]{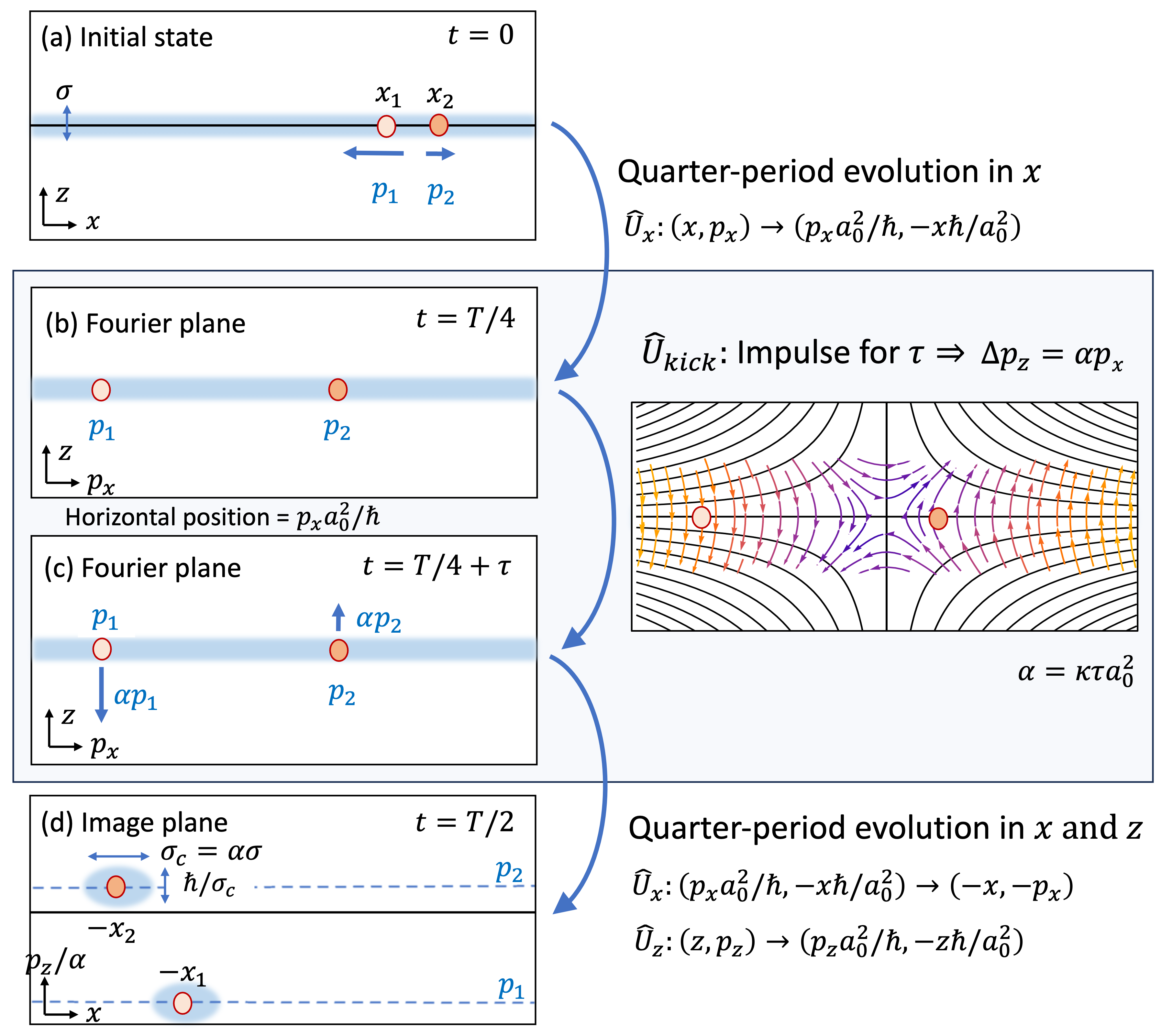}
    \caption{
    Sketch of the protocol for mapping momentum to an auxiliary dimension to realize a 1D Husimi-Q phase-space microscope. (a) Two atoms with initial positions $x_1$ and $x_2$ and momenta $p_1$ and $p_2$ in the physical $x$ dimension. (b) A $T/4$ pulse is applied to move to the Fourier plane {in the $x$ direction}{: the {horizontal} position of a particle now encodes its  {initial} momentum so is labelled by $p_{1,2}$.} 
    {(c)} A spatially dependent impulse maps this momentum to a momentum in the auxiliary $z$ dimension. {(d)} A second $T/4$ pulse along both directions creates an image in the $x$ direction and maps the acquired $z$ momentum to a $z$ displacement. 
    {The protocol gives a choice of how to distribute uncertainty between $x$ and $p_x$.}
    } 
        \label{fig:1}
\end{figure}
The particle starts in the state 
$\Psi_{\rm i}(x,z) = \psi_{\rm i}(x) \phi_0(z)$
where $\psi_{\rm i}(x)$ is the state we wish to characterise. {We take $\phi_0(z)= {\rm e}^{-z^2/2\sigma^2}/(\pi\sigma^2)^{1/4}$, assuming the particle to be in the groundstate of a harmonic confinement in $z$ with oscillator length $\sigma$.} We encode the momentum $p_x$ in $z$ by applying a unitary operator
\begin{equation}
\hat{U}_{\rm kick} = {\rm e}^{{\rm i} \alpha \hat{p}_x \hat{z}/\hbar}
\label{eq:xkick}
\end{equation}
that boosts the momentum in the $z$-direction by  $p_x$ scaled by a  factor $\alpha$. 
To do so, one can use a {pulsed harmonic trap {in $x$} of trapping frequency $\omega=2\pi/T$ for a} quarter-period {duration $T/4$} to map momentum $p_x$ onto position $x$, thereby transforming to the Fourier plane. One then applies a potential $V_{\rm kick}(x,z) = -\hbar \kappa x z$ for a short time $\tau \ll T$, which imparts an $x$-dependent impulse $\hbar \kappa\tau x$ along $z$. Such a saddle potential can be realised by a magnetic field away from the Helmholtz condition\footnote{This can be achieved via the coil geometry or a current imbalance, with the 1D system oriented at 45 degrees to the coil axis.}. Acting with another quarter-period harmonic trap in $x$ then maps back to position space. Denoting the action of each quarter-period harmonic trap (with oscillator length $a_0=\sqrt{\hbar/M\omega}$) by the unitary operator $\hat{U}_x$, the overall transformation is 
\begin{equation}
\hat{U}_x {\rm e}^{{\rm i} \kappa\tau \hat{x} \hat{z} }\hat{U}_x  = \hat{U}_x^2 \;{\rm e}^{{\rm i} \kappa\tau (\hat{p}_x a_0^2/\hbar) \hat{z} }
\end{equation}
which matches the desired form (\ref{eq:xkick}) with $\alpha = \kappa\tau a_0^2$. The prefactor $\hat{U}_x^2$ just reverses the spatial co-ordinate $x\to -x$ and adds an unimportant phase~\footnote{If the second quarter-period oscillator has a different length $a_0'$, the transformation also provides an overall magnification $(a_0'/a_0)^2$.}.
Finally, we measure the  $x$-position and $p_z$-momentum of the particle in this final state, which we denote by $x_{\rm m} \equiv  -x$ (to remove the spatial reversal) and $p_{\rm m} \equiv p_z/\alpha $ (to recover $p_x$ by removing the scaling by $\alpha$). The measurement of $p_{z}$ can be conveniently achieved by applying a quarter-period harmonic oscillator in $z$ simultaneously with the other directions and measuring in position space. {The final positions can be imaged by capturing in a lattice, where for a reasonable matter-wave magnification, the discrete position measurement from the lattice will not be limiting. {The phase space can then be read out via an image of the $x,z$ plane.}

Following this protocol, the probability of measuring the particle in the state $(x_{\rm m},p_{\rm m})$ is readily found to be~\cite{SupMat}
\begin{eqnarray}
P(x_{\rm m}, p_{\rm m})  & = & \left|\langle \psi^{(x_{\rm m},p_{\rm m})}_{\rm coh}| \psi_{\rm i}\rangle\right|^2
\label{eq:olap}
\end{eqnarray}
where
$\psi^{(x_{\rm m},p_{\rm m})}_{\rm coh}(x) \equiv 
{\rm e}^{-(x-x_{\rm m})^2/2\sigma_c^2 + {\rm i}p_{\rm m}x/\hbar}/(\pi \sigma_{\rm c}^2)^{1/4} $
is the phase-space coherent state centred on $(x_{\rm m},p_{\rm m})$ with positional uncertainty $\Delta x=\sigma_c = \alpha\sigma$. Eqn.~(\ref{eq:olap}) defines the Husimi-Q representation of the quantum state: {it is related to the Wigner function by a convolution that renders it a positive-definite measure in phase space}~\cite{leonhardt2010essential}. Eqn.~(\ref{eq:olap}) shows that our protocol is a POVM in the coherent state basis~\cite{Peres1995}. Indeed, it implements precisely the Arthurs-Kelly proposal for measurement of  non-commuting variables~\cite{Arthurs1965}. 
Similar POVMs using two non-commuting spin observables were previously measured in  trapped ions~\cite{Rowe2001} and for collective observables~\cite{Kunkel2019}, while Wigner functions of individual atoms were previously obtained by motional state tomography in tight harmonic traps~\cite{Kanem2005}\cite{Brown2023}. Our work goes beyond these by being applicable to systems of very general form.

{Joint measurements of conjugate variables~\cite{Arthurs1965} leave freedom to choose how to distribute quantum noise between the two variables while remaining consistent with the uncertainty principle. In our protocol,
the uncertainty of the measured momentum is set by the size $\sigma$ of $\phi_0(z)$ and the scaling $\alpha$ in (\ref{eq:xkick}) to be $\Delta p_{\rm m} \sim \hbar/(\alpha\sigma)$.}  
{Reducing this must cause the measurement to add noise} to $\Delta x_{\rm m}$~\cite{Arthurs1965}. 
{Here this arises} because the potential $V_{\rm kick}(x,z)$, applied in the Fourier plane, imparts an impulse $\hbar \kappa\tau z$ in $x$ {due to the necessarily bent force lines}, with uncertainty $\sim \hbar \kappa\tau\sigma$ due to the spread $\sigma$ in $z$. Moving back to real space using $\hat{U}_x$, this introduces the noise $\Delta x_{\rm m} \sim (\hbar\kappa\tau\sigma) (a_0^2/\hbar) =  \alpha \sigma$.

One can readily generalise this procedure~\cite{SupMat},  extending the system of interest to 2D ${\bm r} = (x,y)$, and
imprinting a general
$\hat{U}^{G}_{\rm kick} = {\rm e}^{{\rm i}  G(\hat{\bm p}) \hat{z}}$
where $G({\bm p})$ is a chosen scalar function of ${\bm p} = (p_x,p_y)$.
This generalisation requires an isotropic harmonic oscillator with equal frequencies in $x$ and $y$, as well as imaging in a three-dimensional quantum gas microscope.
Note also that the full phase space is four-dimensional, so cannot be fully encoded in these three spatial dimensions with the scalar function $G({\bm p})$.
One can instead use an internal spin degree of freedom as an additional auxiliary coordinate, which allows the mapping of both components of the 2D momentum onto the Bloch sphere, analogous to a stereographic projection (see Fig.~\ref{fig:2} and~\cite{SupMat}.).
Finally, although we have focused on a single particle, the method can be extended to multi-particle systems of bosons or fermions, provided interactions are sufficiently weak during the quarter-period evolutions that scattering can be neglected. It also admits extensions to spinful systems, provided internal levels can be separately measured.

We refer to the above protocol as a {\it ``Husimi-Q phase-space microscope''}, since it measures the wavefunction in phase space according to Eq.~(\ref{eq:olap}).
We use this term broadly, to include the situations where, jointly with the atomic positions, the microscope images only a subset of the momentum components or some function of momentum.

The Husimi-Q framework also reveals a natural path to a second class of measurement. The joint position--momentum measurement has position uncertainty $\sigma_c$ and momentum uncertainty $\hbar/\sigma_c$. As $\sigma_c\to 0$, and position becomes sharply resolved, one might expect that momentum information is lost. However, this is not the case. Instead, even if one cannot  resolve the full distribution of momentum, one can still measure phase-space averages of the momentum density or its moments.  For example, in the above 1D Husimi-Q microscope, if instead of measuring the momentum in the $z$ direction, one measures the probability to occupy the $n$-th harmonic oscillator eigenstate, one finds this to be proportional to the $2n$-th moment of the local momentum distribution, even as $\sigma_c\to 0$. The resulting {\it ``averaged-mode phase-space microscopes''} are functions only of the spatial co-ordinates and are not subject to Heisenberg noise.

Averages in phase space are readily obtained when one makes the auxiliary space finite: either as a finite set of states of a harmonic trap or as a spin of finite length. Consider the auxiliary space to be an internal spin-$S$ with the system initially prepared in the maximally polarised spin state, \textit{i.e.},  $|\Psi_{\rm i} \rangle = |\psi_{\rm i}\rangle \otimes |S,S\rangle$. We encode information about the system state $|\psi_{\rm i}\rangle$ in the spin using a momentum-dependent spin rotation,  $\hat{U}_{\rm rot} = {\rm exp}\left[-{\rm i} \vec{\theta}(\hat{{\bm p}})\cdot\hat{\vec{S}}\right]$. One can again do this by transforming to the Fourier plane  by quarter-period evolutions in both  $x$ and $y$ (via the product of operators $\hat{U}_x\hat{U}_y$), and then imposing a spatially dependent spin rotation $\vec{\theta}({\bm r})$, before transforming back to  position space with  $\hat{U}_x\hat{U}_y$. {Here we provide two example protocols to measure different moments of the momentum distribution.}

\begin{figure}

\begin{tikzpicture}
\node[anchor=south east,inner sep=0] () at (-0.6,0) {\includegraphics[width=0.27\textwidth]{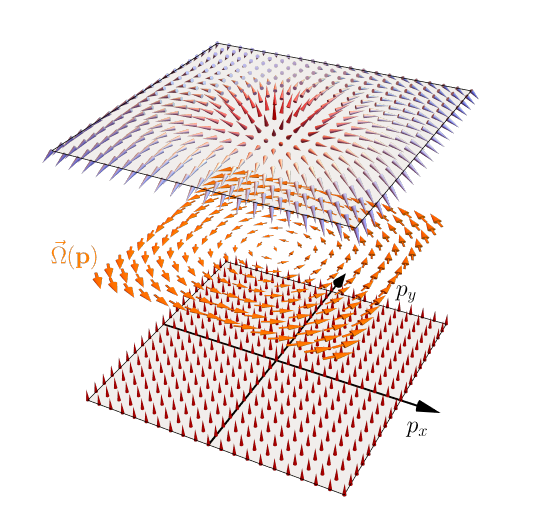}};
\node[anchor=south west,inner sep=0] () at (-1,0) {\includegraphics[width=0.27\textwidth]{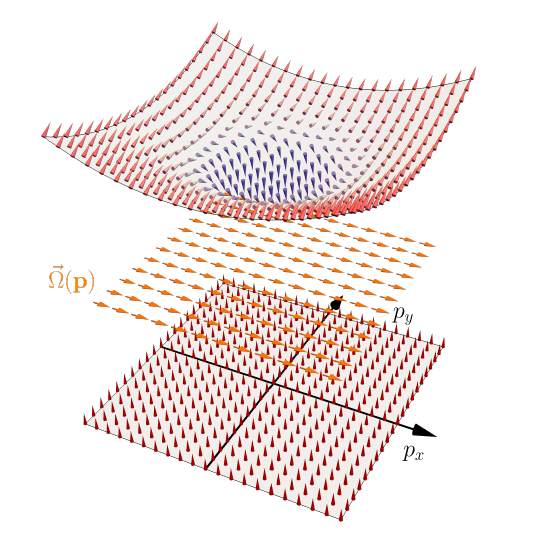}};

\node[align=center,black] at (-4.8,4.3) {(a)};
\node[align=center,black] at (-0,4.3) {(b)};

\end{tikzpicture}

    \caption{Sketches of protocols to map moments of the momentum density to an auxiliary spin state in averaged-mode phase-space microscopes. {These are applied in the Fourier plane, so the position is labelled by momentum ${\bm p}$.} (a) Spin-rotation-(i) transforms the initial $|S,S\rangle$ spin state (aligned vectors of lower layer) using a resonant Rabi pulse that circulates as $\vec{\Omega}\propto (-p_y,p_x,0)$ (orange central layer) to map the 2D momentum space onto the Bloch sphere in the final state (vectors of upper layer), in a mapping analogous to a stereographic projection (see End Matter). (b) Spin-rotation-(ii) uses a uniform Rabi pulse $\vec{\Omega}\propto (1,0,0)$ in the presence of a {quadratically varying Zeeman shift} to imprint  $|{\bm p}|^2$ in $S_z$. {In both cases, the measured spin encodes moments of the momentum distribution.}}
    \label{fig:2}
\end{figure}

{\it Spin-rotation-(i)} We map the 2D momentum to the surface of the Bloch sphere, by a spin rotation  $\vec{\theta}({\bm p})= 2(\,\mp p_y,p_x,0)/p_0$, where $p_0$ sets a characteristic momentum scale~\footnote{At small momenta, this is an approximation to the stereographic projection discussed in the End Matter.}.
In the {matter-wave} Fourier plane (where ${\bm p}$ is transcribed to ${\bm r}$) this requires a resonant Rabi field $\vec{\Omega}({\bm p})\propto \vec{\theta}({\bm p})$ that is circulating in the plane, Fig.~\ref{fig:2}(a). Such a spin rotation could be performed via an optical two-photon transition using Gauss-Laguerre beams with angular momenta $L=0$ and $L = \pm 1$~\footnote{Despite the near-detuning required, the losses should be small, as one only needs to make a small fractional rotation of the levels, as in Ref.~\cite{Cheuk2012}.}.
Overall this {couples} space and spin as
 \begin{equation}
  |\Psi_{\rm f}\rangle =\hat{U}_y^2\hat{U}_x^2\,
  {\rm e}^{-2{\rm i}\frac{(\hat{p}_x \hat{S}_y \,\mp \hat{p}_y \hat{S}_x)}{p_0}}|\psi_{\rm i}\rangle \otimes |S,S\rangle
  \label{eq:spin-i}
  \,,\end{equation}
where  $\hat{U}_y^2\hat{U}_x^2$ effects the inversion ${\bm r}\to -{\bm r}$. This protocol can, in principle, be used to construct a  POVM for the full 4-dimensional phase space of $(x,y)$ and $(p_x,p_y)$~\cite{SupMat}}.

Here we show that projective measurements of the spin components extract moments of the momentum density.
Consider projective measurements of $\hat{S}_z$. Expanding the exponential (\ref{eq:spin-i}) for small $|\hat{\bm p}|/p_0$ gives
$|\Psi_{\rm f}\rangle \propto\hat{U}_y^2\hat{U}_x^2 \sum_{j=0}^{2S}\sqrt{\begin{pmatrix} 2S \\ j \end{pmatrix}}\left(\frac{\hat{p}_x\pm{\rm i}\hat{p}_y}{p_0}\right)^j |\psi_{\rm i}\rangle \otimes |S,S-j\rangle\,$. The particle will be found in the $|S,S-1\rangle $ state at  position ${\bm r}_{\rm m} = -{\bm r}$ 
(chosen to correct for the spatial inversion)   with probability density proportional to $|\langle {\bm r}_{\rm m}|(\hat{p}_x\pm{\rm i}\hat{p}_y)|\psi_{\rm i}\rangle|^2 \equiv 2M[{\cal K} \mp \hbar {\cal M}$], where
\begin{eqnarray}
    {\cal K}({\bm r}_{\rm m}) & \equiv & \frac{\hbar^2}{2M} \left|{\bm \nabla} \psi_{\rm i}({\bm r}_{\rm m})\right|^2  
    \label{eq:kewf} \\
    {\cal M}({\bm r}_{\rm m})  & \equiv &  \frac{\hbar}{2{\rm i}M} {\bm \nabla}\psi_{\rm i}^*\times {\bm \nabla}\psi_{\rm i} \cdot\hat{{\bm z}}
\label{eq:magwf}
\end{eqnarray}
are the kinetic energy density~\footnote{The kinetic energy density can be written as $(-\hbar^2/2M)\psi({\bm r}^*) \nabla^2\psi({\bm r})$ or $(+\hbar^2/2M) |\nabla \psi({\bm r})|^2 $, which gives the same total kinetic energy under integration when $\psi$ vanishes on the boundary.} and the orbital magnetization (the curl of the current density). Thus, repeating measurements for $\Delta L = \pm 1$ allows both the kinetic energy and orbital magnetization densities to be measured. Similarly, the particle will be in the $|S,S-j\rangle $ state with probability density 
$\propto
|\langle {\bm r}_{\rm m}|(\hat{p}_x\pm {\rm i}\hat{p}_y)^j|\psi_{\rm i}\rangle|^2$.
{Thus, the measurement of all spin states provides $2S$ moments of the momentum distribution as well as the particle density ($j=0$)}. 
Measurements of $\hat{S}_{x,y}$ provide the local momentum density~\cite{SupMat}.

{\it Spin-rotation-(ii)} In place of the circulating Rabi field one can engineer a useful momentum-dependent coupling using RF pulses. We impose a {quadratically varying Zeeman shift} in the Fourier plane,  as realised in a magnetic trap (Fig.~\ref{fig:2}b), leading to a detuning $\delta \equiv \Omega|{\bm p}|^2/p_0^2$ when the RF is resonant at $|{\bm p}|^2=0$. We apply a {\it uniform} Rabi field  $\Omega \hat{S}_x$ for a time $\pi/\Omega$, which rotates a particle at $|{\bm p}|^2=0$ into $|S,-S\rangle$, in general in a multiphoton transition.
A particle at $|{\bm p}|^2\neq 0$ will not perform an exact $\pi$ rotation, leading to 
\begin{equation}
|\Psi_{\rm f} \rangle = 
\hat{U}_y^2\hat{U}_x^2
  \exp[\pi {\rm i} ( \hat{S}_x - \hat{S}_z{|\hat{\bm p}|^2}/p_0^2 )]|\psi_{\rm i}\rangle \otimes |S,S\rangle \,.\label{eq:zeemansqrot} \end{equation}
Expanding in ${|\hat{\bm p}|^2}/p_0^2$ shows that the amplitude to be in $|S,-S+j\rangle $ grows as $|\hat{\bm p}|^{2j}$. In particular, the particle will be in the $|S,-S+1\rangle $ state at position ${\bm r}_{\rm m} = -{\bm r}$   with probability proportional to
\begin{equation}
|\langle  {\bm r}_{\rm m}||{\hat{\bm p}}|^2|\psi_{\rm i}\rangle|^2 = \left|\hbar^2 \nabla^2 \psi_{\rm i}({\bm r}_{\rm m})\right|^2 \,,
    \label{eq:kesqwf}
\end{equation}
providing a quartic moment of the momentum density, for $|{\bm p}|\ll p_0$.

Thus, measurements of $\hat{S}_z$ return spatial maps of  moments of the momentum density. Larger values of $S$ allow access to higher moments.  However, $S=1/2$  is already sufficient to measure the kinetic energy density and orbital magnetization (\ref{eq:kewf}) with spin-rotation-(i), or the quartic moment of momentum (\ref{eq:kesqwf}) with spin-rotation-(ii). 
As mentioned above, there is no fundamental limit to the spatial resolution of these averaged-mode measurements, since they do not measure separate non-commuting operators. Rather, they measure certain momentum-dependent observables.
For an $N$-particle system,  the observables Eqns.~(\ref{eq:kewf},\ref{eq:magwf},\ref{eq:kesqwf}) correspond to 
\begin{eqnarray}
    \hat{\rho}_{\rm KE}({\bm r}) & \equiv &  \sum_{i=1}^N \sum_{\alpha=x,y}\frac{1}{2M} \hat{ p}_i^\alpha \delta({\bm r}-\hat{{\bm r}}_i)\hat{ p}_i^\alpha\,,
    \label{eq:kedensity}\\
     \hat{\rho}_{\rm mag}({\bm r}) & \equiv & {\bm \nabla} \times \sum_{i=1}^N \frac{1}{2}\left[ \frac{\hat{\bm p}_i}{M}, \delta({\bm r}-\hat{{\bm r}}_i)\right]_+ \cdot{\hat{\bm z}\,,}    \label{eq:magdensity}\\
\hat{\rho}_{\rm quartic}({\bm r}) & \equiv & \sum_{i=1}^N  |\hat{\bm p}_i|^2 \delta({\bm r}-\hat{{\bm r}}_i)|\hat{\bm p}_i|^2\,,\label{eq:quarticdensity}
\end{eqnarray}
where $[\hat{A},\hat{B}]_+$ denotes the anticommutator of $\hat{A}$ and $\hat{B}$. veraged-mode phase-space microscopes  
allow measurements of these new observables and their correlations, in addition to the particle density $    \hat{\rho}({\bm r})\equiv \sum_{i=1}^N \delta({\bm r}-\hat{{\bm r}}_i)$. In effect, by making the derivatives of the wavefunction observables, they provide continuum realisations of protocols for measuring off-diagonal coherences on lattices, which can measure local currents and kinetic energies~\cite{Impertro2024,PhysRevX.13.011049,kiser2025}. The freedom to control the evolution in the Fourier plane provides a very rich set of possible momentum-based observables to be realised using averaged-mode phase-space microscopes.

In practice, the spatial resolution of the measurement process will be limited by atomic shot noise. For a 2D system with mean density $\bar{\rho}$, measurements of density $\hat{\rho}({\bm r})$ with spatial resolution $w$ have an error $\delta \rho / \bar{\rho} \sim 1/\sqrt{\bar{\rho} w^2}$ {due to the projection noise from the atom number in that area.} Similarly, measurements of the kinetic energy density $\hat{\rho}_{\rm KE}({\bm r})$ for particles with characteristic wavelength $\lambda$ 
will have an error $\delta \rho_{\rm KE} / \overline{\rho_{\rm KE}} \sim 1/\sqrt{(\bar{\rho}/\lambda^2)(\hbar^2/p_0^2)w^2}$, because the typical density of spin-flips is $(\bar{\rho}/\lambda^2)(\hbar^2/p_0^2)$. Thus, to have a signal with minimal noise, one should choose $p_0 \sim \hbar/\lambda$. Note that the above protocols (\ref{eq:spin-i},\ref{eq:zeemansqrot}) approximate the densities (\ref{eq:kedensity},\ref{eq:quarticdensity}) accurately for $|{\bm p}|\ll p_0$, with
$p_0$ acting as a high-momentum cutoff.

{The proposed phase-space microscopes could be directly implemented in state-of-the-art quantum gas microscopes. They would significantly expand the toolbox of cold-atom experiments:
by exposing momentum-space features with spatial resolution, 
they open up many new ways to probe quantum many-body systems.
Here we focus on weakly trapped gases where the continuum momentum is the relevant quantity. However, the microscope can also be used for lattice-based systems, where it can be helpful to use band-mapping to recover the lattice quasimomentum~\cite{YangWeitenbergCooper_prep}. In the following, we present four  applications in continuum settings.

\noindent {\it (1) Short-range external potentials.} 
Imagine that one  wishes to characterise a sharp potential step imposed at the edge of a system, leading to a wavefunction  $\psi_{\rm edge}(x)$ 
that vanishes over a small lengthscale $\delta$. Measurement of the particle density  $|\psi_{\rm edge}(x)|^2$ with a point-spread function in position of $w$ 
can resolve down to $\delta\sim w$.  However,  a  Husimi-Q phase space microscope with the same positional resolution $\sigma_c = w$, {evaluated at} the edge can detect a high momentum tail out to $\sim \hbar/\delta$  even for $\delta\ll w$ (See Fig.~\ref{fig:edge}).
While a global measurement of the momentum distribution would also depend on the sharpness of the edge, the phase-space measurement targets the local properties and can lead to a larger signal-to-noise. 

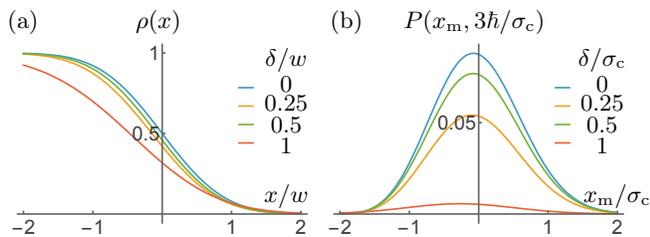
\begin{figure}
    \centering

\begin{tikzpicture}
\node[anchor=south east,inner sep=0] () at (0,0) {\includegraphics[width=0.22\textwidth]{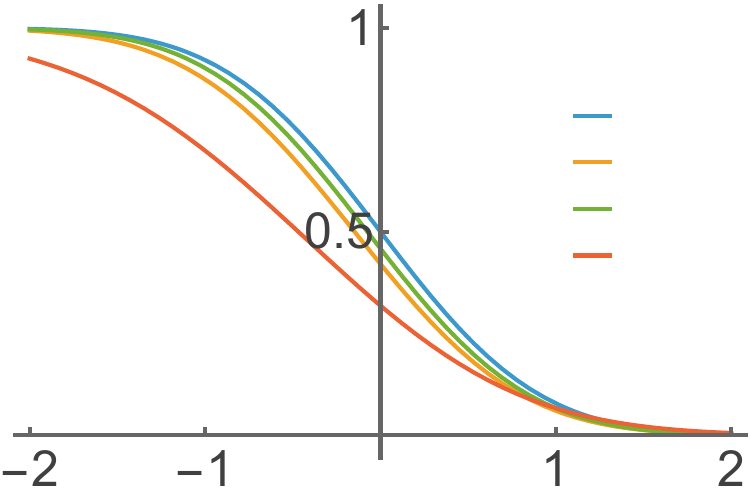}};
\node[anchor=south west,inner sep=0] () at (0.25,0) {\includegraphics[width=0.22\textwidth]{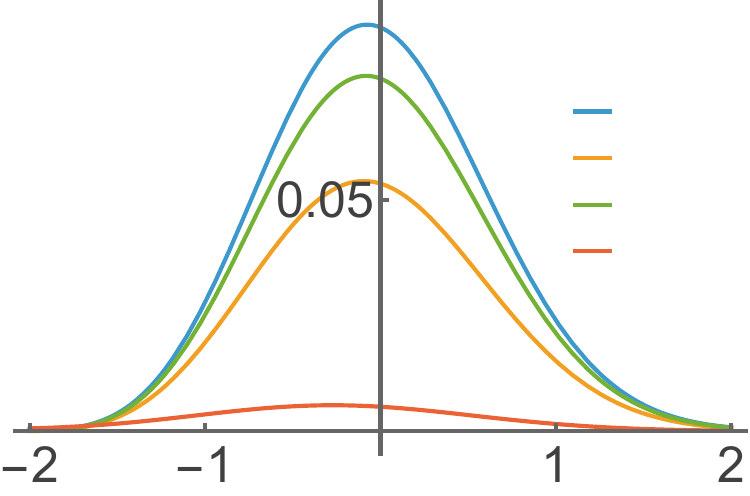}};

\node[align=center,black] at (4.1,0.58) {$x_{\rm m}/\sigma_{\rm c}$};
\node[align=center,black] at (-0.3,0.58) {$x/w$};

\node[align=center,black] at (-3.8,2.9) {(a)};
\node[align=center,black] at (0.5,2.9) {(b)};

\node[align=center,black] at (2.2,2.9) {{\small $P(x_{\rm m},3\hbar/\sigma_{\rm c})$}};

\node[align=center,black] at (3.9,2.4) {$\delta/\sigma_{\rm c}$};
\node[align=center,black] at (3.9,2.07) {{\small $0$}};
\node[align=center,black] at (3.9,1.82) {{\small $0.25$}};
\node[align=center,black] at (3.9,1.53) {{\small $0.5$}};
\node[align=center,black] at (3.9,1.26) {{\small $1$}};

\node[align=center,black] at (-2,2.9) {{\small $\rho(x)$}};
\node[align=center,black] at (-0.3,2.4) {$\delta/w$};
\node[align=center,black] at (-0.3,2.07) {{\small $0$}};
\node[align=center,black] at (-0.3,1.82) {{\small $0.25$}};
\node[align=center,black] at (-0.3,1.53) {{\small $0.5$}};
\node[align=center,black] at (-0.3,1.26) {{\small $1$}};

\end{tikzpicture}

    \caption{Measurement of a sharp step in the potential, with edge thickness $\delta$  mimicked by a state $\psi_{\rm edge}(x) = [1-\tanh(x/\delta)]/2$.   (a) Density distribution as measured by a conventional QGM with point spread function $\propto {\rm e}^{-x^2/w^2}$. (b) 
    Distribution of particles with momentum $p_{\rm m}=3\hbar/\sigma_{\rm c}$ as measured by a Husimi-Q phase space microscope with  $\sigma_c = w$.
      The density measurement (a) becomes insensitive to $\delta$ for $\delta/w\lesssim 0.5$, while the phase-space measurement (b) continues to discriminate via the high-momentum tail.}
        \label{fig:edge}
\end{figure}

\noindent {\it (2) Imaging quantum vortices}.  Averaged-mode phase space microscopes allow access to the velocity field associated with the circulation around a vortex. They can be used to image the kinetic energy and magnetization densities (\ref{eq:kedensity},\ref{eq:magdensity}) which have peaks at the vortex core, or the quartic density (\ref{eq:quarticdensity}) , which is a ring around the vortex core,  Fig.~\ref{fig:vortex}. 
These features have sizes set by the healing length $\xi$, so sensitivity is maximized for $p_0\sim \hbar/\xi$.
The measurements could also be performed in 3D, using spin as the auxiliary space. {Then the paths of the vortices will appear as lines of spin-flips along their cores.}
This may facilitate measurements of quantum turbulence~\cite{Navon2016}. }

\begin{figure}
    \centering

\begin{tikzpicture}
\node[anchor=south west,inner sep=0] () at (0,0) {\includegraphics[width=0.4\textwidth]{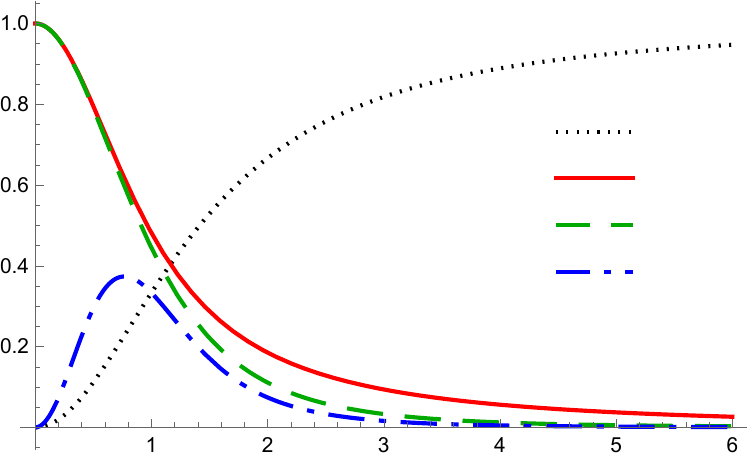}};
\node[align=center,black] at (6.75,3.15) {$\frac{\langle \hat{\rho}(r)\rangle } {\bar{\rho}}$};

\node[align=center,black] at (6.75,2.7) {$\frac{\langle \hat{\rho}_{\rm KE}(r)\rangle }{{\bar{\rho}}\hbar^2/\xi^2}$};

\node[align=center,black] at (6.75,2.2) {$\frac{\langle \hat{\rho}_{\rm mag}(r)\rangle }{{\bar{\rho}}\hbar/\xi^2}$};

\node[align=center,black] at (6.8,1.75) {$\frac{\langle \hat{\rho}_{\rm quartic}(r)\rangle }{{\bar{\rho}}\hbar^4/\xi^4}$};

\node[align=center,black] at (7.5,0.5) {$r/\xi$};
\end{tikzpicture}

    \caption{Particle density, kinetic energy density, magnetization density, and quartic momentum density around a 2D vortex of healing length $\xi$, computed for the ansatz~\cite{FetterRMP} $\psi(r,\theta) =\sqrt{\bar{\rho}} \,r\, {\rm e}^{{\rm i}\theta} /\sqrt{r^2+2\xi^2}$. {The averaged-mode phase-space microscope also allows measurements of the latter three quantities.}     }
        \label{fig:vortex}
\end{figure}
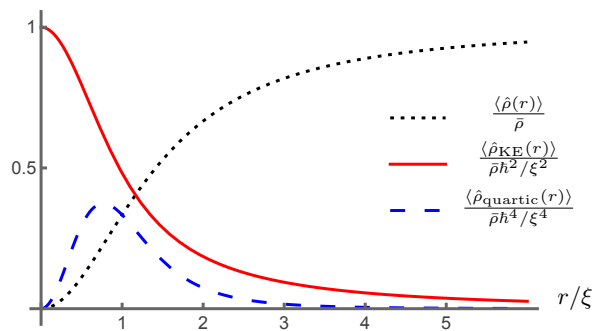

\noindent {\it (3) Thermometry.} Measurements of the local kinetic energy allow access to the local temperature. This could facilitate thermometry of strongly interacting systems -- \textit{e.g.}, by partitioning the cloud into a region of interest and a region in which the equation of state is well known. For non-interacting bosons in 2D, a thermal cloud with density $\bar{\rho}$ has a kinetic energy density 
$\bar{\rho}_{\rm KE}= k_{\rm B}T \bar{\rho}$, while a BEC has $\bar{\rho}_{\rm KE}=\frac{\pi M}{12 \hbar^2} (k_{\rm B}T)^2$, both providing simple scales. (See~\cite{SupMat}  for other cases.)
The protocol would also allow a better fit to bimodal profiles in a partially condensed cloud by including spatial position and kinetic energy for each particle. Such precise determination of the condensate fraction is, \textit{e.g.}, important for studying thermal condensate fluctuations~\cite{Kristensen2019,Christensen2021,Jalali-Mola2025}.
The local nature also opens up the possibility to measure temperature gradients, or indeed to measure second sound.

\noindent {\it (4)  Local Tan Contact.}
Quite generally, for a gas with short-range interactions, the high-momentum population takes the form $\rho_{\bm p} \sim C\hbar^2/|{\bm p}|^{4}$.  Here $C$ is the Tan contact, which characterizes the interacting system~\cite{TAN20082971}. Depending on spatial dimension and on which moment of the momentum is measured, this can dominate the signal in an averaged mode phase space microscope.   For example, for 2D systems the local quartic density Eq.~(\ref{eq:quarticdensity}) is dominated by this contribution when $p_0 \gtrsim \hbar\sqrt{\bar\rho}$~\footnote{For a weakly interacting 2D BEC, $C\sim ({\tilde g} \bar{\rho})^2$ so it contributes $\langle \hat{\rho}_{\rm quartic}\rangle \sim {\tilde g}^2 {\bar \rho}^2 p_0^2/\hbar^2$. For $\bar{\rho}\,p_{0}^2/\hbar^2\gg 1$,  this is larger than the contribution from the condensate $\langle \hat{\rho}_{\rm quartic}\rangle \sim \bar\rho (\hbar/\xi)^4$ expected in the vicinity of a vortex or edge, with healing length $\xi\sim 1/\sqrt{{\tilde g}\bar{\rho}}$.}. Using the protocol (\ref{eq:zeemansqrot}) for $S=1/2$, one finds  $\langle \hat{\rho}_{\rm quartic}\rangle  = 0.14 \hbar^2 Cp_0^2$. While $C$ has been measured by other methods~\cite{Stewart2010,PhysRevLett.105.070402,zou2021tan,huang2025direct}, the averaged-mode phase-space microscope permits measurements even in spatially inhomogeneous systems, \textit{e.g.}, probing its variation in vortex cores even for
strongly interacting superfluids. More generally, the dependence of $\langle \hat{\rho}_{\rm quartic}\rangle$ on $p_0$ would allow a detailed analysis of condensate depletion in such inhomogeneous settings.

Beyond the applications outlined above, phase space microscopes offer even more powerful ways to probe strong correlated phases, through measurements of {spatial} correlations of momentum-space features. In the lowest Landau level regime, correlations of density and velocity have been shown to provide a way to measure the many-body interaction energy~\cite{Spasic-Mlacak2025}.
We expect that correlations in phase-space also have the potential to reveal hidden order in exotic many-body phases, potentially in combination with machine learning analysis of the data~\cite{Carleo2019}.
The access to conjugate observables with detection of every particle might be useful for accessing entanglement structure~\cite{Islam2015} similar to the measurement of non-commuting spin observables~\cite{Kunkel2022}. Squeezing between particles in a quantum many-body system could directly show in the phase-space microscope via a narrower distribution in one of the conjugate variables of position and momentum.

%{\bf Data Availability Statement}
%The data that support the plots presented in this paper and other findings of this study are available from the corresponding author upon reasonable request. The authors declare no competing financial interests.

\acknowledgments{{We thank Ulrich Schneider and Zoran Hadzibabic for helpful comments on the manuscript.}
This work was funded by EPSRC (Grant Nos. EP/V062654/1 and EP/Y01510X/1), by a Simons Investigator Award (Grant No. 511029) and by the Deutsche Forschungsgemeinschaft (DFG, German Research Foundation) via Research Unit FOR 5688, Project No. 521530974.}

%\bibliography{ref}


\begin{widetext}
    \newpage

\end{widetext}

\section{End Matter}

\subsection{General Analysis for Auxiliary Spatial Co-ordinate}

\label{app:generalz}

We consider a particle in an initial state $\psi_{\rm i}({\bm r})$ within the plane described by ${\bm r} = (x,y)$ and a state $\phi_0(z)$ along the perpendicular direction $z$. We shall encode the momentum ${\bm p}=(p_x,p_y)$ in the $z$-momentum by applying a generalized kick
\begin{equation}
\hat{U}^{G}_{\rm kick} = {\rm e}^{{\rm i} G(\hat{\bm p}) \hat{z}}
\label{eq:gkick}
\end{equation}
where $G({\bm p})$ is a scalar function.

We write the initial state of the particle
\begin{equation}|\Psi_{\rm i}\rangle = \int {\rm d}^2{\bm r}\, {\rm d}z\, \Psi_{\rm i}({\bm r},z) |{\bm r},z\rangle \end{equation}
with
$\Psi_{\rm i}({\bm r},z) = \psi_{\rm i}({\bm r}) \phi(z) $.
We retain (weak) confinement in $z$ and apply a harmonic trap to both $x$ and $y$ for a $T/4$ pulse. We denote the oscillator length of this trap by $a_0$, such that this pulse transforms the wavefunction via the operator
\begin{equation}
\hat{U}_{x} = {\rm exp}{[-{\rm i} (\hat{p}_x^2a_0^2/\hbar^2+\hat{x}^2/a_0^2)\pi/2]}\,.
\end{equation}
Using the fact that 
$\hat{U}_{x} = \sum_n |
n\rangle \langle n|\, {\rm exp}{[-{\rm i} (n+1/2)\pi/2]}$
and the fact that the oscillator states $|n\rangle$  take the same functional form in both position and momentum space, up to scaling, one finds that
the application of quarter period pulses to both $x$ and $y$ produces the ``Fourier plane'' for the ${\bm r} = (x,y)$ co-ordinates, where 
$|\Psi_{\rm Fourier}\rangle = \hat{U}_{x}\hat{U}_{y}|\Psi_{\rm i}\rangle $. 
In this Fourier plane, the position ${\bm r}$  encodes the momentum ${\bm p}$ of the initial state.
Formally this arises from $\hat{U}_{x}^{-1} \hat{x} \hat{U}_{x} = -\hat{p}_x a_0^2/\hbar$ and $\hat{U}_{x}^{-1} \hat{p}_x \hat{U}_{x} = \hat{x}\hbar/a_0^2$.

We now apply a kick in the $z$ direction with a strength that depends on the position ${\bm r}$. To match the desired form (\ref{eq:gkick}) this is
\begin{equation}
    |\Psi'_{\rm Fourier}\rangle = \exp[-{\rm i}G(\hat{\bm r}\hbar/a_0^2) \hat{z}] \hat{U}_{x}\hat{U}_{y}|\Psi_{\rm initial}\rangle  \,.
\end{equation}
We transform back from the Fourier plane to real space by applying the $T/4$ evolution in $x,y$ again. Then  
\begin{eqnarray}
|\Psi_{\rm f}\rangle & = & \hat{U}_{x}  \hat{U}_{y}\exp[{\rm i} G(\hat{\bm r}\hbar/a_0^2) \hat{z}] \hat{U}_{x} \hat{U}_{y}
|\Psi_{\rm initial}\rangle\\
& = & - \hat{R}_{x}  \hat{R}_{y} \exp[{\rm i} \hat{z} G(\hat{{\bm p}})] |\Psi_{\rm initial}\rangle
\end{eqnarray}
where we have used $\hat{U}_x^2 = (-{\rm i})\hat{R}_x$ with ${\hat R}_x$ the reflection operator, taking $|x\rangle \to |-x\rangle$.

Finally we measure the particle to be at position ${\bm r}_{\rm f}$ in the $xy$ plane and in the momentum state $p_{z,{\rm f}}$ along $z$. (The measurement of momentum in $z$ is conveniently performed by converting $p_z$ to position $z_{\rm f}$ via $\hat{U}_z$, and then measuring $z_{\rm f}$.)
Introducing a complete sets of states in the $xy$ plane, the probability to detect the particle at ${\bm r}_{\rm f}$ and $p_{z,{\rm f}}$ is
\begin{widetext}
    \begin{equation}
 | \langle {\bm r}_{\rm f}; p_{z,{\rm f}}|\Psi_{\rm f}\rangle|^2 =\frac{1}{(2\pi\hbar)^{3}}\left| \int {\rm d}^2 {\bm p}\,  {\rm d}^2 {\bm r}\, {\rm d} z\,
{\rm e}^{-{\rm i}{\bm p}\cdot({\bm r}+{\bm r}_{\rm f})/\hbar}
{\rm e}^{{\rm i}[G({\bm p})-p_{z,{\rm f}}/\hbar]z}
\Psi_{\rm initial}({\bm r},z)\right|^2\,.
\label{eq:final}
\end{equation}
This general expression (\ref{eq:final}) allows one to analyse the consequences of different kicks $G$ in the Fourier plane.
The example used in the main text is the 1D phase space $(x,p)$ with $G(p) = \alpha p/\hbar$. 
The final co-ordinates $(x_{\rm f},p_{z,{\rm f}})$ used in (\ref{eq:final}) are expressed in terms of
$x_{\rm m} \equiv - x_{\rm f}$ and $p_{\rm m} = p_{z,{\rm f}}/\alpha$ to correct for the spatial inversion and the scaling of momentum with $\alpha$.
Using the 1D version of Eqn.~(\ref{eq:final}) this leads to the probability
\begin{eqnarray}
P(x_{\rm m}, p_{\rm m}) & = & \frac{\alpha}{(2\pi\hbar)^2}\left|
  \int {\rm d}p\, {\rm d}x\, {\rm d}z\, {\rm e}^{{\rm i} (z\alpha + x_{\rm m}-x)p/\hbar-{\rm i}\alpha p_{\rm m}z/\hbar}\psi_{\rm i}(x)\phi_0(z)\right|^2
=\left|
  \int {\rm d}x\, {\rm e}^{-{\rm i} (x-x_{\rm m})p_{\rm m}/\hbar}\frac{{\rm e}^{-\frac{(x-x_{\rm m})}{2\alpha^2\sigma^2}}}{(\pi \alpha^2\sigma^2)^{1/4}}\right|^2\,.
\end{eqnarray}

\end{widetext}

{For 2D systems, one can envisage encoding any scalar function of ${\bm p}$ in the transverse momentum, for example using $G({\bm p}) \propto |{\bm p}|^2$. The final imaging could be performed via a spatial tomography consisting of successive images of different planes~\cite{Eliasson2020,Koepsell2020} or other techniques such as helical point-spread function engineering~\cite{Legrand2024}.}

\subsection{General Analysis for Auxiliary Spin-$S$}
\label{app:generalspin}

The spin-rotation (i) used in the main text (\ref{eq:spin-i}) is an approximation to the stereographic projection of the 2D momentum to {} spin coherent states} on the Bloch sphere. 

We consider an internal spin-$S$. A stereographic projection that maps from the 2D momentum $(p_x,p_y)$ to the surface of a sphere $(n_x,n_y,n_z)$ with $|\vec{n}|^2=1$, is described by
\begin{equation}
    \frac{n_x+{\rm i}n_y}{1-n_z} = \frac{p_x+{\rm i}p_y}{p_0}
\end{equation}
where $p_0$ is a characteristic momentum scale, such that $|{\bm p}| = p_0$ is mapped to the equator, $n_z=0$. 
The key step in the mapping is to perform a unitary transformation that rotates the initial fully spin-aligned state $|S,+S\rangle$ to the spin-coherent state that is aligned with this $\vec{n}$. 
This is achieved by choosing a rotation $\vec{\theta}({\bm p})$ such that 
  \begin{equation} 
\hat{U}_{\rm rot} =   \exp[2{\rm i} \hat{\vec{S}}\cdot (\hat{p}_y,-\hat{p}_x,0)/{|\hat{{\bm p}}|}\arctan(|\hat{\bm p}|/p_0) ]\,.\end{equation}
  A  state of definite momentum $(p_x,p_y)$ is mapped to the coherent spin state
$|\chi_{\bm p}\rangle \propto \hat{U}_{\rm rot}({\bm p}) |S,S\rangle$.
For small momenta $|{\bm p}|\ll p_0$, we have $\arctan(|\hat{\bm p}|/p_0)\simeq |\hat{\bm p}|/p_0$ to get 
  \begin{equation} |\chi_{\bm p}\rangle  \simeq 
  \exp[2{\rm i} \hat{{\bm S}}\cdot {\bm e}_z\times (\hat{p}_{y},-\hat{p}_x,0)/p_0) ]|S,S\rangle\,,\end{equation}
which is the form used in the main text.

The spin coherent states are~\cite{Radcliffe_1971}
\begin{equation}
    |\eta\rangle = \frac{1}{(1+|\eta|^2)^S}\exp{(\eta \hat{S}_-)}|S,S\rangle\,,
\end{equation}
with $\eta\equiv \eta_x+{\rm i}\eta_y$ a complex number. Like the phase-space coherent states, the spin coherent states provide an overcomplete basis.
As such, one can imagine performing POVMs in the $|\eta\rangle$ basis. To do so would require the spin-$S$ to be mapped onto another (higher dimensional) Hilbert space, with a spin with $S'\gg S$, in the spirit of Ref.~\cite{Kunkel2019}. 

Imagine that one were to perform the POVM in the spin-coherent states. The outcome would be to measure the particle at position ${\bm r}_{\rm f}$ and spin-coherent state $|\eta_{\rm f}\rangle$ with probability $P({\bm r}_{\rm f},\eta_{\rm f}) \equiv |\langle {\bm r}_{\rm f},\eta_{\rm f}| {\Psi}_{\rm f}\rangle  |^2$ where
    \begin{eqnarray}
    \langle {\bm r}_{\rm f},\eta_{\rm f}| {\Psi}_{\rm f}\rangle  
 & \propto & \int {\rm d}^2 {\bm p}\,  {\rm d}^2 {\bm r} \,
{\rm e}^{-{\rm i}{\bm p}\cdot({\bm r}+{\bm r}_{\rm f})}
\Psi_{\rm i}({\bm r})\langle \eta_{\rm f} | \chi_{\bm p}\rangle \,.
\label{eq:finalspincoherent}
\end{eqnarray}
Using 
\begin{equation}
    \langle \eta_{\rm f}| \chi_{\bm p}\rangle = \frac{(1+\eta_{\rm f}^*\chi_{\bm p})^{2S}}{(1+|\eta_{\rm f}|^2)^S(1+|\chi_{\bm p}|^2)^S}
\end{equation}
and assuming $|\eta_{\rm f}|,|\chi_{\bm p}|\ll 1$, we obtain
\begin{equation}\langle \eta_{\rm f}| \chi_{\bm p}\rangle \simeq \exp[-S(|\eta_{\rm f}|^2+|\chi_{\bm p}|^2-2\eta_{\rm f}^*\chi_{\bm p})] 
\label{eq:fspin}
\end{equation}
with  $\chi_{\bm p} \simeq (p_x+{\rm i}p_y)/p_0$. The integral over momentum in (\ref{eq:finalspincoherent}) can then be done, to give
\begin{equation}
   P({\bm r}_{\rm f},\eta_{\rm f}) =  \left|\langle {\bm r}_{\rm f},\eta_{\rm f}| {\Psi}_{\rm f}\rangle \right|^2  = \left|\langle \psi_{\rm coh}^{({\bm r}_{\rm f},\eta_{\rm f})}|\psi_{\rm i}\rangle\right|^2
\end{equation} where the position representation of the state $|\psi_{\rm coh}^{({\bm r}_{\rm f},\eta_{\rm f})}\rangle$ is
\begin{widetext}
\begin{equation}
    \langle {\bm r}|\psi_{\rm coh}^{({\bm r}_{\rm f},\eta_{\rm f})}\rangle = \frac{p_0}{\sqrt{2\pi S}\hbar}{\rm e}^{{\rm i}{( \eta_{{\rm f},x},\eta_{{\rm f},y})}\cdot ({\bm r}-{\bm r}_{\rm f})p_0/\hbar}{\rm e}^{-\frac{1}{4S}\{[(x-x_{\rm f})p_0/\hbar+2S\eta_{{\rm f},y}]^2+
(y-y_{\rm f})p_0/\hbar-2S\eta_{{\rm f},x}]^2\}}\,.
\end{equation}
\end{widetext}
This is a coherent state of the {\it two-dimensional} phase space, centred on the position $(x_{\rm f}-2S\eta_{{\rm f},y}\hbar/p_0,y_{\rm f}+2S\eta_{{\rm f},x}\hbar/p_0)$ and momentum $(p_x,p_y) = p_0(\eta_{{\rm f},x},\eta_{{\rm f},y})$. Thus the measurement is a POVM in the full 2D phase space, and would realize a 2D Husimi-Q phase space microscope.  The positional uncertainty is controlled by $\sigma_{\rm c}' \equiv \sqrt{2S}\hbar/p_0$ and the momentum uncertainty by $\hbar/\sigma_{\rm c}'$. For fixed characteristic momentum $p_0$, one achieves increasingly accurate resolution in momentum by increasing the size of the spin $S$. 
A proper choice of the atomic species would allow large spin manifolds such as 17 states in Dysprosium~\cite{Chalopin2020}
or 23 states in Holmium~\cite{Saffman2008}.
For Dy ($S=8$) this gives $\sigma_{\rm c}' = 4\hbar/p_0$ and for Ho ($S=11$) $\sigma_{\rm c}' \approx 4.7\hbar/p_0$.

Even without full POVMs on the spin-$S$, this mapping of momentum to spin can be readily used to construct an averaged-mode phase space microscope using standard projective measurements of the spin components $\hat{S}_{x,y,z}$.
Since $|\chi_{\bm p}\rangle$ is the spin coherent state aligned along $\vec{n}\simeq \left(p_x/p_0,p_y/p_0, 1- |{\bm p}|^2/(2p_0^2)\right)$ one has
\begin{equation}
    \langle \chi_{\bm p}| \hat{\vec{S}}|\chi_{\bm p}\rangle \simeq S\left(p_x/p_0,p_y/p_0, 1- |{\bm p}|^2/(2p_0^2)\right)\,.
\end{equation}
Measurements of the expectation values of $\hat{S}_x$ and $\hat{S}_y$ provide information on the mean momentum densities, while the expectation value of $\hat{S}_z$ gives the mean kinetic energy density as discussed in the main text.

\subsection{Thermodynamics of the Weakly Interacting BEC}
\label{app:thermodynamics}
For a weakly interacting Bose gas in 2D, the low-temperature thermodynamics is dominated by the Bogoliubov sound mode with $E_{\bm p} = s|{\bm p}|$.   The speed of sound is $s= \sqrt{{\tilde g}\bar{\rho}}(\hbar/M)$ with ${\tilde g}$ the dimensionless coupling constant, which is ${\tilde g} =  \sqrt{8\pi a_s/\ell_z} $ for atoms of scattering length $a_s$ confined to 2D by a oscillator state of size $\ell_z$. 
Applying Bogoliubov theory one finds
\begin{equation}
    \bar{\rho}_{\rm KE}(T) = \bar{\rho}_{\rm KE}(0) + \frac{\zeta(3)}{2\pi \hbar^2s^2 }(k_BT)^3\,,
\end{equation}
with $\zeta(3) \simeq 1.202$. The zero-temperature value $\bar{\rho}_{\rm KE}(0)$ has a divergent contribution $\propto C \ln(p_{\rm max})$,
arising from the contact~\cite{TAN20082971}, which sets the large momentum tail of the momentum distribution, $\rho_{p} \sim C\hbar^2/p^{4}$. The momentum cut-off $p_{\rm max}$ is set by non-universal physical effects: the ramping times in the experimental dynamics~\cite{Stewart2010}; or, as discussed in the main text, the characteristic scale $p_0$ up to which the quadratic expansion of the observable is valid.  This generates a non-universal background value for $\bar{\rho}_{\rm KE}$. However, the temperature can be extracted from the change $\Delta \bar{\rho}_{\rm KE} = \bar{\rho}_{\rm KE}(T)-\bar{\rho}_{\rm KE}(0)$ with $T\propto (\Delta \bar{\rho}_{\rm KE})^{1/3}$.

\end{document}